# Spatially Continuous and High-resolution Land Surface Temperature: A Review of Reconstruction and Spatiotemporal Fusion Techniques

**Penghai Wu[a,b,*], Zhixiang Yin[a], Chao Zeng[c], Sibo Duan[d], Frank-Michael Göttsche[e], Xiaoshaung Ma[a], Xinghua Li[f], Hui Yang[b], Huanfeng Shen[c,*]**

[a] Anhui Province Key Laboratory of Wetland Ecosystem Protection and Restoration, Anhui University, Hefei, Anhui 230601, China;

[b] Institute of Physical Science and Information Technology, Anhui University, Hefei, Anhui 230601, China;

[c] School of Resource and Environmental Science, Wuhan University, Wuhan, 430079, China;

[d] Institute of Agricultural Resources and Regional Planning, Chinese Academy of Agricultural Sciences, Beijing 100081, China;

[e] Institute of Meteorology and Climate Research (IMK), Karlsruhe Institute of Technology (KIT), Karlsruhe, Germany;

[f] School of Remote Sensing and Information Engineering, Wuhan University, Wuhan, 430079, China

## ABSTRACT

Remotely sensed, spatially continuous and high spatiotemporal resolution (hereafter referred to as high resolution) land surface temperature (LST) is a key parameter for studying the thermal environment and has important applications in many fields. However, difficult atmospheric conditions, sensor malfunctioning and scanning gaps between orbits frequently introduce spatial discontinuities into satellite-retri1eved LST products. For a single sensor, there is also a trade-off between temporal and spatial resolution and, therefore, it is impossible to obtain high temporal and spatial resolution simultaneously. In recent years the reconstruction and spatiotemporal fusion of LST products have become active research topics that aim at overcoming this limitation. They are two of most investigated approaches in thermal remote sensing and attract increasing attention, which has resulted in a number of different algorithms. However, to the best of our knowledge, currently no review exists that expatiates and summarizes the available LST reconstruction and spatiotemporal fusion methods and algorithms. This paper introduces the principles and theories behind LST reconstruction and spatiotemporal fusion and provides an overview of the published research and algorithms. We summarized three kinds of reconstruction methods for missing pixels (spatial, temporal and spatiotemporal methods), two kinds of reconstruction methods for cloudy pixels (Satellite Passive Microwave (PMW)-based and Surface Energy Balance (SEB)-based methods) and three kinds of spatiotemporal fusion methods (weighted function-based, unmixing-based and hybrid methods). The review concludes by summarizing validation methods and by identifying some promising future research directions for generating spatially continuous and high resolution LST products.

* Corresponding author:

**Penghai Wu** (wuph@ahu.edu.cn). Tel.: +86 18715012409.

Anhui Province Key Laboratory of Wetland Ecosystem Protection and Restoration, Anhui University, Hefei, Anhui 230601, China;

**Huanfeng Shen** (shenhf@whu.edu.cn). Tel.: +86 13163235536.

School of Resource and Environmental Science, Wuhan University, Wuhan, 430079, China.

## 1. Introduction

Land surface temperature (LST) is a crucial parameter in investigating environmental and ecological processes (Hansen et al. 2010; Tierney et al. 2008), and is also valuable in studies of evapotranspiration, soil moisture conditions, heat-related health issues and urban heat islands(Anderson et al. 2012; Sellers et al. 1997). From a climate perspective, LST is important for evaluating land surface and land-atmosphere exchange processes, constraining surface energy budgets and model parameters, and providing observations of surface temperature change both globally and in key regions(Guillevic et al. 2017). Satellite remote sensing offers the only possibility to measure LST over extended regions with acceptable temporal resolution and complete spatial coverage(Li et al. 2013b; Wan et al. 2004).

Satellite-derived thermal infrared (TIR) data has a relatively high spatial resolution with acceptable accuracy(Wan et al. 2004). Various algorithms (e.g., single-channel, split-window, and temperature and emissivity separation) have been devised to derive operational LST products(Li et al. 2013b). For example, the Moderate Resolution Imaging Spectroradiometer (MODIS)(Wan et al. 2004), the FengYun-2/3 (FY-2/3) Visible Infrared Scanning Radiometer (VIRR) and the Spinning Enhanced Visible and InfraRed Imager (SEVIRI) LST products (Trigo et al. 2008) are available for public use. However, TIR-based retrieval algorithms only work well for data acquired under clear-sky conditions and without any instrument faults(Duan et al. 2017). The spatial continuity of LST is strongly affected by pixel with invalid or missing values caused by clouds or cloud shadow (hereafter referred to as cloudy pixels). On average cloudy-sky conditions account for more than half of the day-to-day weather around the globe(Jin 2000). For example, more than 60% of MODIS LST are contaminated by clouds(Cornette and Shanks 1993). Furthermore, some cloud-free but naturally bright pixels are frequently classified as cloud covered and their corresponding LST are set to the missing pixel value (Yang et al. 2019). For another, for a single sensor, the trade-off between temporal and spatial resolution often makes it impossible to obtain LST with the high temporal and spatial resolutions required by some applications(Wu et al. 2015c). Generally, LST retrieved from sensors with fine spatial resolution have poor temporal resolution, which leads to temporal discontinuities, as shown in Figure 1(a).

Satellite passive microwave (PMW) measurements are attractive for retrieving (sub-)surface temperature, especially under cloudy conditions, because they are much less affected by clouds and water vapor than TIR measurements(Holmes et al. 2016; Shwetha and Kumar 2016). However, the spatial resolution of PMW measurements (e.g., 25 km for AMSR-E) is much lower than that of TIR measurements. Besides, missing information caused by defective sensors (e.g., Landsat ETM+ SLC-off data)(Shen et al. 2016a) and scanning gap between orbits (e.g., Auqa/AMSR-E、GCOM/AMSR2 data) also introduce spatial discontinuities into LST products (Duan et al. 2017).

The above-mentioned spatial discontinuities and the restrictions on simultaneous spatial and temporal resolution seriously hinder applications of LST products in many fields. For instance, urban heat islands (UHI) can be continuously observed with AMSR-E, FY-2/3 and MODIS etc., but their associated spatial resolutions are too coarse to reveal detailed UHI spatial patterns. More spatial details can be observed in Landsat, ASTER and HJ images; however, due to their long revisit cycles (more than 15 days) and frequent cloud contamination, different LST scenes acquired by these sensors differ considerably in their space-time observation conditions (Shen et al. 2016a). As a result, for practical applications there is an increasingly urgent demand for spatially continuous high-resolution LST products (SCHR-LST).

Because of the growing number of available satellite LST products, many different approaches for generating SCHR-LST have been proposed, resulting in numerous publications on SCHR-LST algorithms and methods. Therefore, it is important and timely to present an overview of the state of the art in SCHR-LST methods. Although there have been earlier a review on the disaggregation of LST to finer temporal and spatial resolutions by Zhan et al. (2013), it reviewed from the perspective of thermal sharpening and temperature unmixing. However, some recently proposed methods (e.g., spatiotemporal fusion methods) to obtain LST with finer temporal and spatial resolutions, also should give a survey. Furthermore, to our best knowledge, a thorough review of methods and algorithms for deriving spatially continuous LST has not been performed.

The objective of this paper is to review methods for generating spatially continuous high-resolution LST, describe the state-of-the-art, and identify the most promising research fields, thereby ultimately benefiting LST producers and developers of SCHR-LST algorithms. The remainder of this paper is structured as follows: section 2 describes the reasons and consequences of spatial discontinuities in satellite LST data and the trade-off between spatial and temporal resolution. Section 3 and Section 4 provide an overview of current reconstruction and spatiotemporal fusion algorithms for generating SCHR-LST. Section 5 presents validation methods for SCHR-LST products. In Section 6, we put forward some prospects for the future study of LST reconstruction and spatiotemporal fusion. Finally, concluding remarks are given in Section 7.

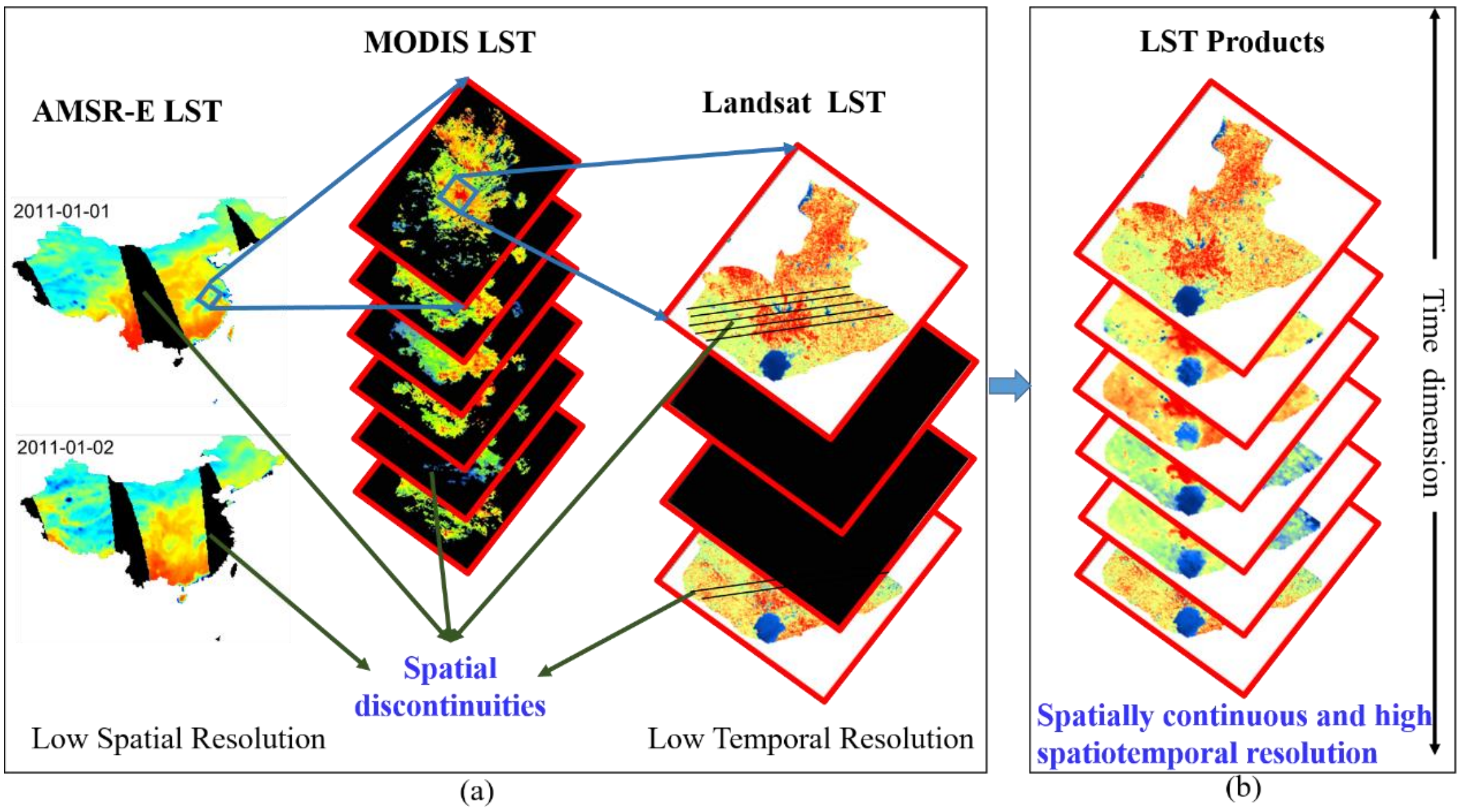

Figure 1. Comparison of spatially discontinuous LST data and low spatial and temporal resolution LST data (a) and an example of spatially continuous LST at high spatiotemporal resolution (b).

## 2. Background

### 2.1 Current satellite-based LST products

Over the past decades, LST estimation from satellite TIR measurements has significantly improved(Li et al. 2013b) and resulted in LST products with acceptable accuracy. Furthermore, many application fields have a growing need for remote sensing data products. Operational LST products are currently retrieved from several instruments and are freely available from various organizations. Some of the most commonly used operational and standard LST products and their specifications are listed in Table 1. Operational LST products are mainly obtained from the Advanced Spaceborne Thermal Emission and Reflection radiometer (ASTER) onboard the Terra satellite (https://lpdaac.usgs.gov/products/ast_08v003/), the Moderate Resolution Imaging Spectroradiometer (MODIS) onboard the Terra and Aqua satellites (https://ladsweb.modaps.eosdis.nasa.gov/search/), the Visible and InfraRed Radiometer (VIRR) onboard the FengYun-3A/B satellites and the Visible Infrared Spin Scan Radiometer (VISSR) onboard the FengYun-2F/G satellites (http://satellite.nsmc.org.cn/portalsite/default.aspx), the Advanced Baseline Imager (ABI) onboard the Geostationary Operational Environmental Satellite (GOES)-R (https://www.avl.class.noaa.gov/saa/products/welcome), the Spinning Enhanced Visible & Infrared Imager (SEVIRI) onboard the Meteosat satellite series (https://landsaf.ipma.pt/en/), the Sea and Land Surface Temperature Radiometer (SLSTR) onboard the Sentinel-3 A/B satellites (http://www.esa.int/Our_Activities/Observing_the_Earth/Copernicus/Sentinel-3/Data_products), and the Visible Infrared Imaging Radiometer Suite (VIIRS) onboard the S-NPP satellite (https://www.avl.class.noaa.gov/saa/products/welcome). Furthermore, some scholars provide open source software for generating LST products, e.g. for Landsat LST (Isaya Ndossi and Avdan 2016). Other LST products used in various publications are only available on request from the corresponding authors, e.g. (Holmes et al. 2016).

Table 1. Common LST satellite products and some of their specifications.

| LST data | Spatial resolution | Temporal resolution | Temporal Extent |
|---|---|---|---|
| AUQA AMSR-E | 25 km | 1 d | 2002/06- 2011/09 |
| GCOM AMSR2 | 10 km | 1 d | 2012/07- Present |
| FY-2F/2G VISSR | 5 km | 1 h | 2012/11- Present |
| GOES Imager | 4 km | 15-30 min | 1994/09- Present |
| MSG SEVIRI | 3 km | 15 min | 2005/02- Present |
| GOES ABI | 2 km | 15min | 2018- Present |
| Terra/Aqua MODIS | 1 km and 5km | 1 d | 2000/06- Present |
| FY-3A/3B/3C VIRR | 1 km | 1 d | 2009/08- Present |
| Sentinel-3 SLSTR | 1 km | 1 d | 2016/02- Present |
| S-NPP VIIRS | 750 m | 1 d | 2011/11- Present |
| ASTER TIR | 90 m | 16 d | 2000/03- Present |
| Landsat TM/ETM+/TIRS | ~100 m | 16 d | 1984-Present |

## 2.2 Problem description

### *2.2.1 Reconstruction*

The main reason for spatially discontinuous LST are missing or cloudy pixels, which severely limits many LST applications. Therefore, suitable methods for filling or amending these pixels are required: a possible cost-effective approach is to reconstruct them by utilizing complementary information from different sources(Shen et al. 2015), an approach that aroused great interest in the remote sensing community.

Reconstructing missing or cloudy information in LST data is an ill-posed inverse problem. For the convenience of our readers and in order to unify the terminology, here we provide a brief introduction to the problem, describe the general ideas behind the reconstruction of LST data, and explain the notation and symbols.

In this paper, vectors and matrices are denoted by bold non-italic characters and scalar variables by italic non-bold characters. Reconstructing missing LST data requires that the missing values can be estimated from the existing / remaining valid LST values and complementary spatial, multi-temporal, spatiotemporal and multi-sensor data. As shown in Fig. 2, LST data **L** are given at observation time $t_0$ and $\mathbf{L} \in \mathrm{R}^{m*n}$ ($\mathbf{L}$: $\mathbf{\Omega} \in \mathbf{R}^2 \rightarrow \mathbf{R}^2$), where $\mathbf{\Omega}$ represents the spatial domain and comprises m*n points. We assume that domain $\mathbf{\Omega}$ contains a missing or cloudy pixel region **MC** and a region **EC** with existing and valid LST, i.e., $\mathbf{\Omega} = \mathbf{MC} \cup \mathbf{EC}$ and $\mathbf{MC} \cap \mathbf{EC} = \mathbf{\Phi}$. The goal of reconstruction is to estimate (i.e. reconstruct) an LST value at position ($x_0$, $y_0$) located in **MC** from pixels in **EC**, which may contain data from other observation times and sensors. LST reconstructions have to be based on reasonable assumptions and are required to be visually natural and in agreement with all known thermal properties. Given a missing LST pixel ($x_0$, $y_0$) obtained from a sensor ($s_0$) and acquired at observation time $t_0$, its LST value can be reconstructed pixel as:

$$LST(x_0, y_0, t_0, \mathrm{s}_0) = \sum_{x=1}^{m}\sum_{y=1}^{n}\sum_{t=t_0}^{t_p}\sum_{s=s_0}^{s_q} f(LST(x, y, t, \mathrm{s})) \tag{1}$$

Where ($x_0$, $y_0$)$\in$**MC**, (x, y)$\in$**EC**, and $f(\cdot)$ is a linear or nonlinear function denoting the relationships of all the existing / remaining valid pixels. t is the observation time and s is the sensor, m and n are the rows and columns of the pending reconstructed LST data, p and q are the numbers of the observation time and sensor. Complementary information from the same sensor can only be exploited if $s=s_0$ while spatial information can only be exploited if $t=t_0$. Furthermore, complementary information from different sensors with the same / similar observation times (e.g. $t_0$) can be used to reconstruct cloudy pixels, e.g. by blending TIR LST (e.g. from MODIS) with PMW LST (e.g. from AMSE-R), and have the potential to produce spatially complete LST datasets.

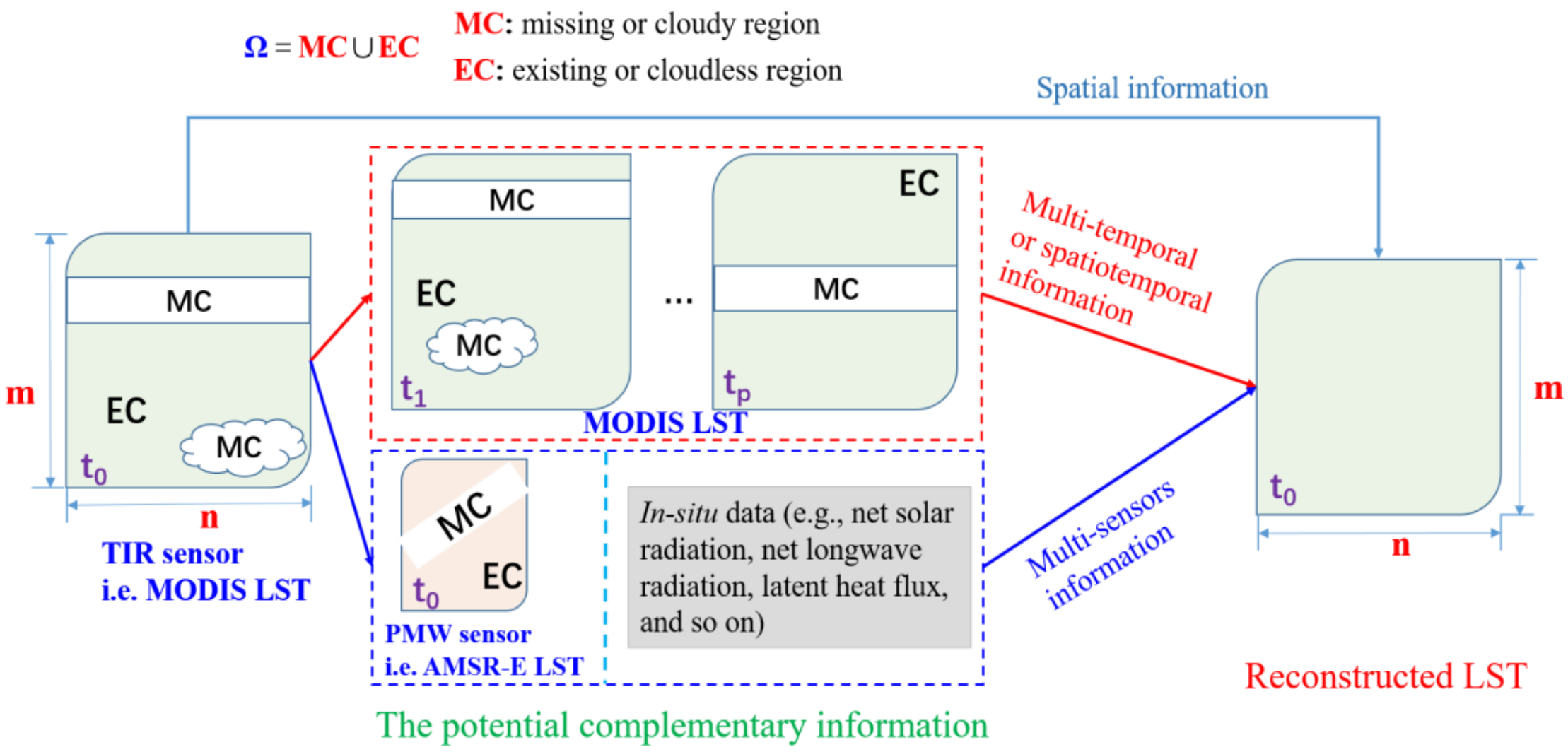

Figure 2. LST reconstruction using complementary information (spatial, multi-temporal, spatiotemporal and multi-sensor).

Figure 2 summarizes two main LST reconstruction methods, i.e. methods using spatial / multi-temporal / spatiotemporal information and methods using multi-sensor information: these two methods are introduced separately in this review. Furthermore, all the reconstruction methods discussed here are exclusively based on satellite LST products, i.e. reconstructions based on thermal infrared radiances are not covered.

### *2.2.2 Spatiotemporal fusion*

Due to technical and financial constraints, there is a trade-off between spatial and temporal resolutions(Gao et al. 2006), i.e. a sensor providing LST data at fine spatial resolution exhibits poor temporal resolution and vice versa(Zhan et al. 2013). For instance, geostationary satellites provide multi-spectral images of the observed Earth disk at frequent time intervals (up to 15 min)(Sun et al. 2006). However, their low spatial resolutions (3–5km) limit the spatial details observed over heterogeneous landscapes (Inamdar et al. 2008). In contrast, polar-orbiting Landsat and MODIS provide LST with spatial resolutions of about 100m to 1000m, respectively, which allows the monitoring of heterogeneous areas in more detail. However, the long revisit-cycle (MODIS: two views per day; Landsat: one view every 16 days) probably miss the optimal observation time, particularly over rapidly changing areas.

Several papers reviewed methods for increasing the spatial and temporal resolution in remote sensing data (Zhan et al. 2013; Zhu et al. 2018b). Different terms are used to refer to the various methods for enhancing LST spatiotemporal resolution, e.g. downscaling (Bechtel et al. 2012; Stathopoulou and Cartalis 2009), image fusion (Quan et al. 2018; Weng et al. 2014; Wu et al. 2015c), and disaggregation(Zhan et al. 2013; Zhan et al. 2016). According to Xia et al, these methods can be classified into two categories: kernel-driven methods, which downscale LSTs via auxiliary data from multi-spectral sensors, and fusion-based methods, which predict fine-resolution LSTs by integrating temporal change and neighborhood information from different sensors(Xia et al. 2019). In their review paper, Zhan et al. focused on methods dedicated to disaggregating land surface temperature and gave a comprehensive and systematic review of kernel-driven methods(Zhan et al. 2013). The last decades witnessed the emergence of various new satellite sensors and LST products: therefore, fusion-based methods have developed rapidly and attracted more and more attention. This review focuses on the development of fusion-based methods over recent decades.

It should be noted that the fusion-based methods presented in this paper are spatiotemporal fusion methods, which differ from more traditional fusion techniques such as spatial-spectral fusion. Such traditional methods generally use a panchromatic band only to enhance the spatial resolution of a multispectral image(Shen et al. 2016b). In contrast, spatiotemporal fusion methods consider different spatial resolutions and acquisition times simultaneously. Furthermore, spatial-spectral fusion is often performed on raw digital numbers, while spatiotemporal fusion generally requires a physical parameter as input, i.e. reflectance or LST. The basic idea behind LST spatiotemporal fusion is to predict fine spatial resolution LST at $t_0$ from coarse spatial resolution LST at the same $t_0$ and a fine spatial resolution scale conversion factor (SCF). The SCF can be obtained from a fine scale classification image or from pairs of fine and coarse spatial resolution LST observed at various times (i.e. $t_1, t_2 ... t_p$). Fine spatial resolution LST (from sensor $s_0$) at time $t_0$ can be modeled as:

$$LST_H(i,t_0) = f\left(LST_L(i,t_0), \mathrm{SCF}\right) \tag{2}$$

where $LST_H$ and $LST_L$ represent the fine spatial resolution LST and the coarse spatial resolution LST resampled to the fine-resolution grid, respectively, and $i$ denotes the i-th pixel.

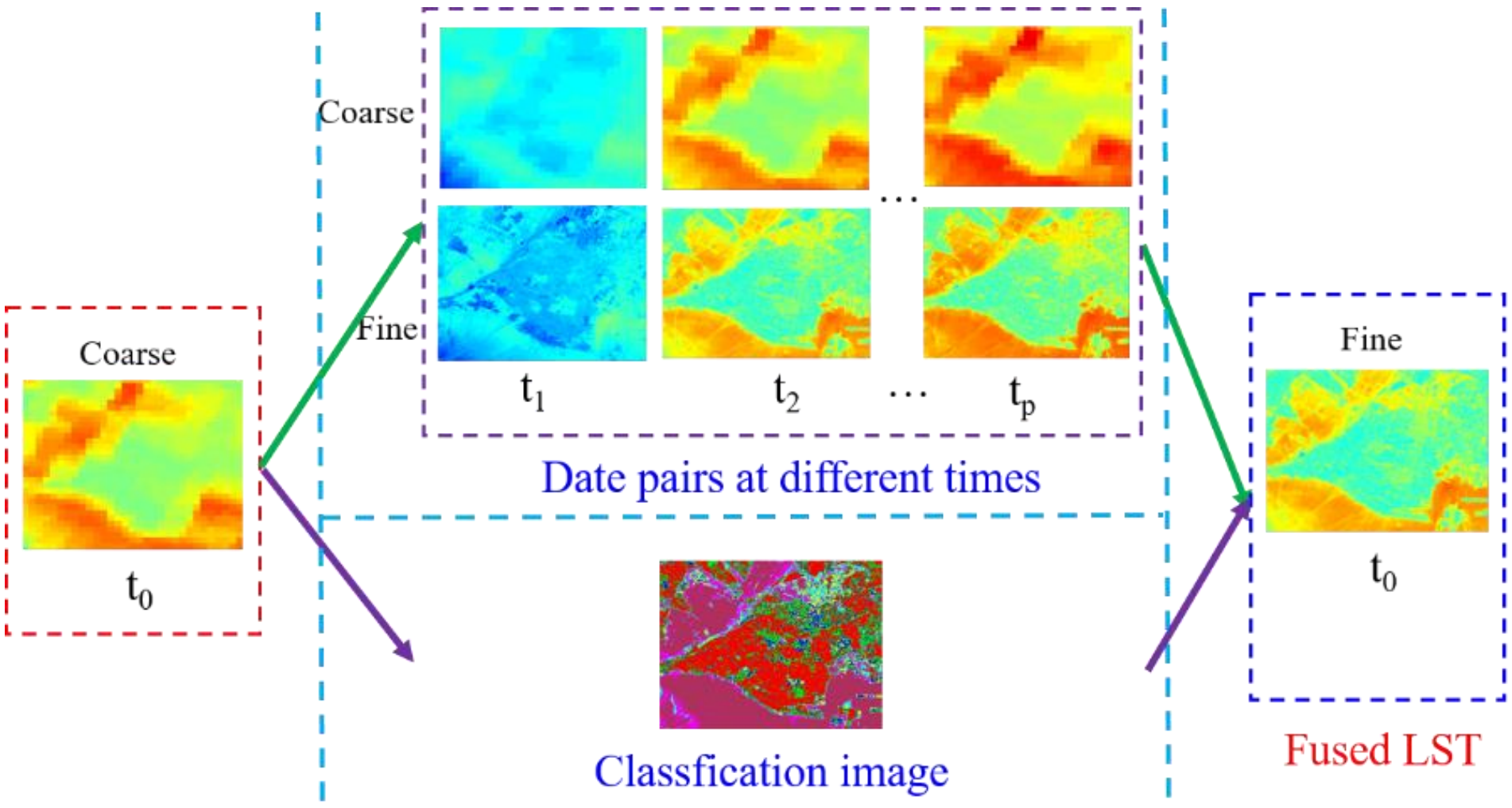

Figure 3. Spatiotemporal LST fusion with a scale conversion factor (SCF) obtained from data pairs (fine and coarse resolution) at different times or a fine scale classification image.

*2.2.3 Connections between reconstruction and spatiotemporal fusion*

Many applications would benefit from spatially continuous LST products on a global scale,

while spatially continuous high-resolution LST products would be ideal for many fine scale applications. Therefore, reconstruction and spatiotemporal fusion methods are often used together, as shown in Figure 4. For instance, Shen et al. employed reconstruction and spatiotemporal fusion methods to analyze the long-term and fine-scale summer surface urban heat island (SUHI) of the city of Wuhan in China(Shen et al. 2016a). Wu et al. compared the diurnal and seasons' SUHI in Hefei, China based on reconstructed and fused Landsat-like LST data(Lu et al. 2018). Spatially complete and temporally continuous LST maps were generated for surface soil moisture mapping by integrating a multi-temporal reconstruction method and a data fusion method (Long et al. 2019). We can extend the basic idea of spatiotemporal fusion by interpreting spatiotemporal fusion as a special case of reconstruction where no auxiliary pixels are available in space at the predicted time (i.e. completely missing data). Based on the low resolution LST information at the predicted time, spatiotemporal fusion can be used to reconstruct remote sensing images affected by phenological changes and significant land cover changes(Shen et al. 2019). Therefore, reconstruction and spatiotemporal fusion are closely interconnected, which is the main reason for reviewing both of them here.

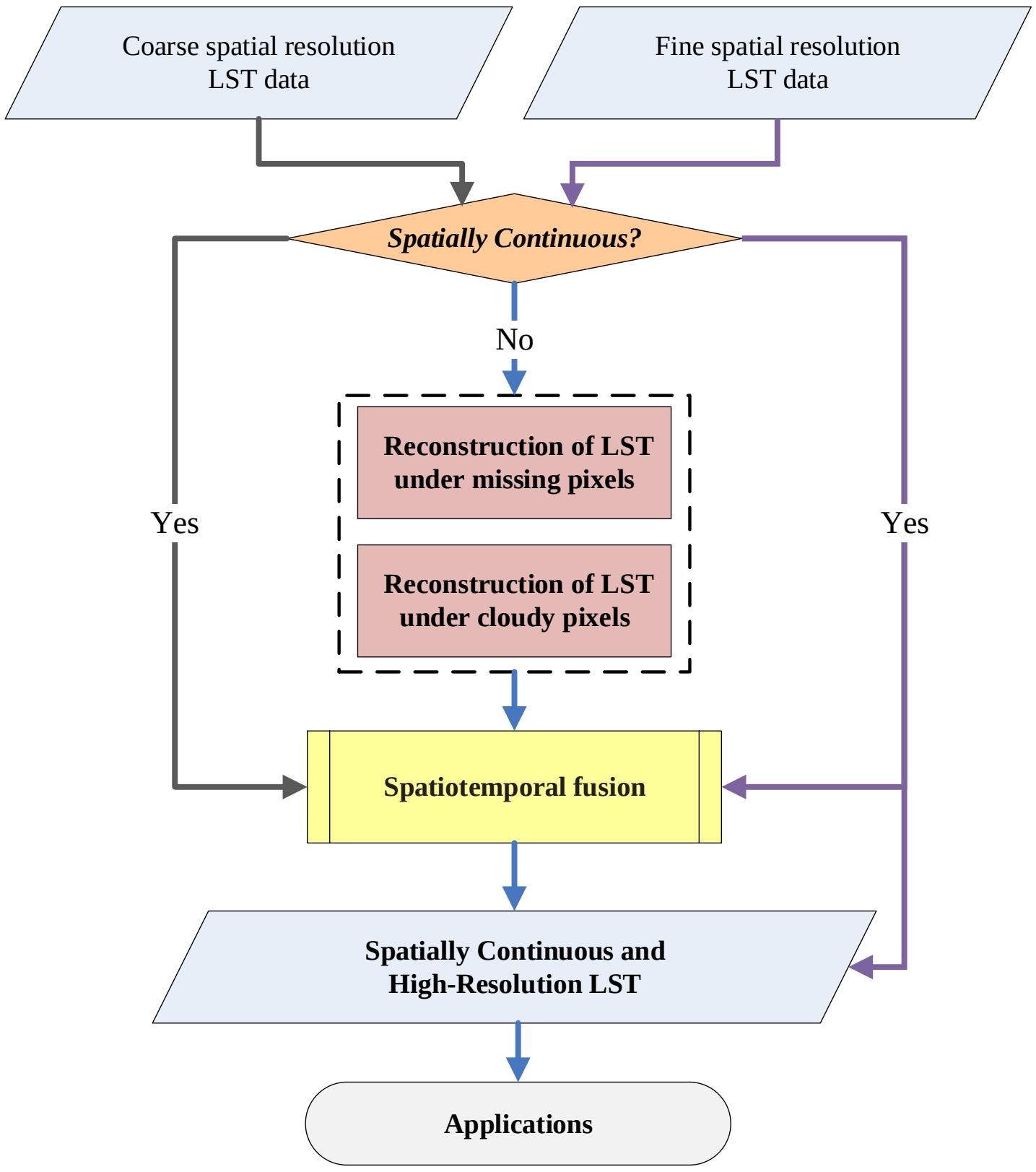

Figure 4. Generation of SCHR-LST and the relationship between LST reconstruction and spatiotemporal fusion.

## 3. Reconstruction for spatial continuity

### 3.1 Reconstruction of missing LST pixels

Although (partially) cloudy pixels are often treated like missing pixels, there are obvious differences when estimating LST for cloudy pixels and for missing pixels from neighboring cloud-free pixels. In this review only those pixels are regarded as missing, for which the measurements are actually missing, i.e. mainly due to defective sensors (e.g. Landsat ETM+ SLC-off data)(Shen et al. 2016a) and scanning gaps between orbits (e.g. Auqa/AMSR-E、GCOM/AMSR2 data), which may occur in TIR and PMW imagery.

Reconstruction techniques can effectively recover missing information and improve the usability of deteriorated LST data and several methods have been developed. Based on the used reference information, the methods can be divided into three categories (Metz et al. 2014): 1) spatial methods, which do not use additional information; 2) multi-temporal methods, which extract complementary information from other data acquired at the same location at different times; 3) spatiotemporal methods, which extract the complementary information from additional spatial and temporal information. An example for multi-temporal reconstruction of missing AMSR-E LST pixels is listed in Figure 5.

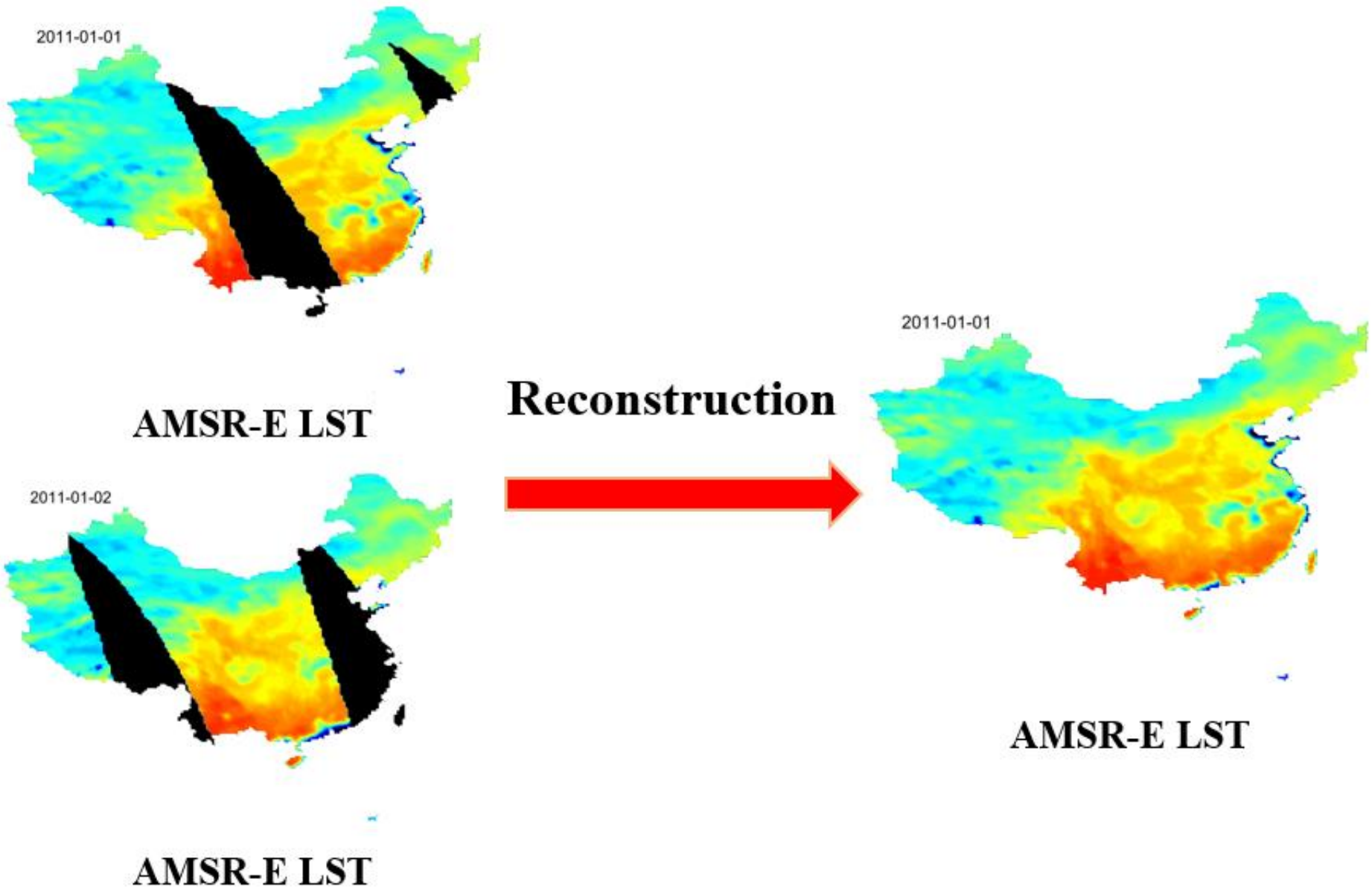

Figure 5. A diagram for multi-temporal reconstruction of missing AMSR-E LST pixels. The target AMSR-E LST images (observed on 2011-01-01) and the multi-temporal AMSR-E LST images (observed on 2011-01-02) are used to reconstruct the missing LST pixels, for generating spatially continuous AMSR-E LST (2011-01-01).

*3.1.1 Spatial reconstruction methods*

Reconstruction methods based exclusively on spatial information are the most traditional of the three approaches. This category of methods reconstruct missing data using the remaining valid LST data. The approach is based on the assumption that the missing data and the remaining data share the same statistical or geometrical structures(Guillemot and Le Meur 2013). The most basic reconstruction methods are spatial interpolation approaches, e.g. inverse distance weighting(Liu et al. 2017), spline function (Kilibarda et al. 2014) and geo-statistical interpolation methods. Motivated by the high spatiotemporal heterogeneity of LST, some studies attempted to take more factors into account and used multi-variate interpolation methods, e.g. cokriging (Ke et al. 2011; Neteler 2010). However, with only limited spatial information available, the resulting LST data over heterogeneous landscapes are often blurred and have unsatisfactory accuracy. Spatial reconstruction methods are generally easy to implement and perform well over homogeneous landscapes with a small number of invalid pixels.

*3.1.2 Multi-temporal reconstruction methods*

Reconstruction methods based on multi-temporal information use temporal images of the same region at different times to reconstruct missing pixels. The algorithms primarily employed are the linear temporal approach(Crosson et al. 2012; Kang et al. 2018; Zeng et al. 2015), the harmonic analysis method(Xu and Shen 2013), the temporal Fourier analysis approach(Scharlemann et al. 2008), the wavelet method(Lu et al. 2007), the asymmetric Gaussian function fitting method, the diurnal temperature cycle (DTC)-based method(Liu et al. 2017; Udahemuka et al. 2008) and dictionary learning(Li et al. 2014b). In contrast to the spatial reconstruction methods, multi-temporal methods often work well for LST images with larger missing regions, but are highly sensitive to temporal inconsistencies, e.g. caused by land cover change or the weather. Generally speaking, multi-temporal methods ignore data from geographically neighboring pixels and are appropriate when differences are mainly linked to regular change, e.g. observation conditions and phenology. In contrast, abrupt changes, e.g. new buildings and man-made landscapes, are difficult to reconstruct (Shen et al. 2015).

*3.1.3 Spatiotemporal reconstruction methods*

Spatiotemporal methods appear to be the most suitable methods to reconstruct missing remote sensing data with high spatiotemporal variability (Metz et al. 2014; Sun et al. 2017). The most basic approach is to sequentially apply a spatial and a temporal reconstruction method, e.g. use a multi-temporal method first and, if this does not recover all missing pixels, apply a spatial reconstruction method to the output. For instance, Liu et al presented a spatiotemporal reconstruction method for missing Feng Yun-2F (FY-2F) LST data. Simulated and real data experiments showed that the method can work well, with root mean square errors (RMSE) of about 2◦ C in most cases (Liu et al. 2017). Duan et al., proposed a spatio-temporal interpolation module for constructing missing AMSR-E temperature data due to orbital gaps between satellite overpasses(Duan et al. 2017). Weiss et al. proposed a gap-filling approach for LST image time-series using neighboring valid data and data from other times, i.e. different calendar dates or multi-annual datasets (Weiss et al. 2014). Although these two-step methods use temporal and spatial information, they do not sufficiently exploit the available simultaneous temporal and spatial information. Maximizing the use of all available temporal and spatial information is the key idea behind spatiotemporal reconstruction methods.

All the aforementioned methods are designed for reconstructing the LST for missing pixels. However, in practice these methods are often used to reconstruct LST under cloudy conditions (Liu et al. 2017; Shen et al. 2015; Sun et al. 2017), even though they can only provide hypothetical clear-sky LST values, which generally differ from LST actually found under cloudy conditions(Liu et al. 2017; Zeng et al. 2018).

### 3.2 Reconstructing LST for cloudy pixels

Cloudy-sky conditions account for about half of the actual day-to-day weather on a global scale and lead to large data gaps in TIR LST imagery(Jin 2000). Therefore, it is highly desirable to devise effective algorithms for reconstructing LSTs under cloudy conditions. A number of techniques and algorithms have been specifically devised and applied to reconstruct cloudy pixel LST (Duan et al. 2017; Fu et al. 2019; Jin 2000; Shwetha and Kumar 2016; Xu et al. 2019; Yang et al. 2019; Yu et al. 2019). These reconstruction techniques generally fall into two categories: the first involves passive microwave (PMW) measurements captured by space-borne sensors (e.g. Advanced Microwave Scanning Radiometer and the Special Sensor Microwave/Imagers), which also retrieve LST under clouds. The second category contains surface energy balance (SEB)-based algorithms that require additional assumptions and/or known meteorological conditions for estimating the LST differences between clear-sky and cloudy pixels. An example for PMW-based reconstruction of cloudy MODIS LST pixels is showed in Figure 6.

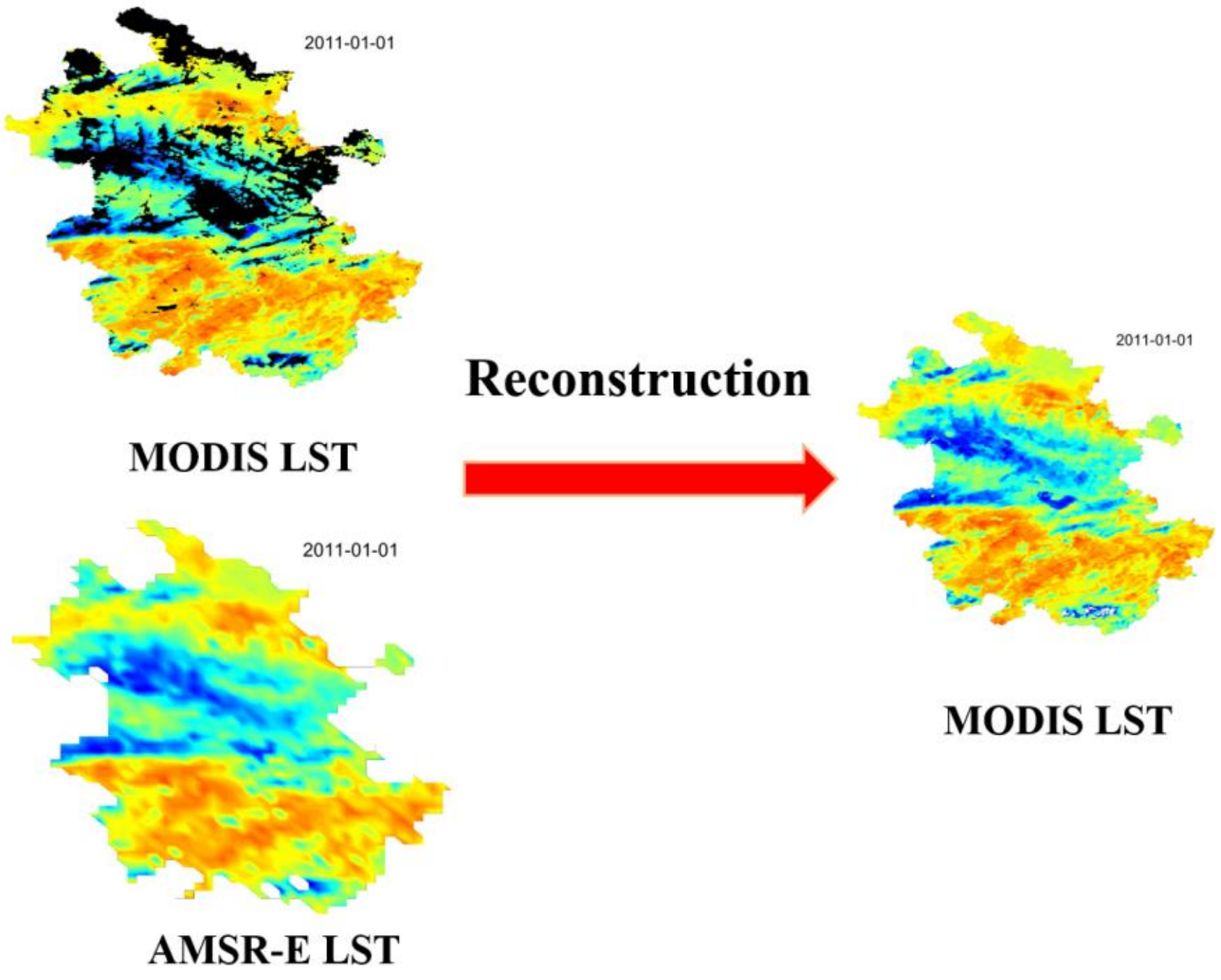

Figure 6. A diagram for PWM-based reconstruction of cloudy MODIS LST pixels. The target MODIS LST images (observed on 2011-01-01) and the AMSR-E LST images (observed on 2011-01-01) are used to reconstruct the cloudy MODIS LST pixels, for generating spatially continuous MODIS LST (2011-01-01).

### *3.2.1 PMW-based methods*

Given the different advantages of TIR and PMW LSTs, methods blending TIR and PMW LST

products have the potential to produce spatially complete LST datasets with high accuracy and moderate spatial resolution(Xu et al. 2019). Recently, several blending methods have been developed for TIR LST and PMW LST. Shwetha and Kumar used artificial neural networks (ANNs) to obtain an LST dataset with spatially complete coverage under cloudy conditions based on the AMSR-E (AMSR2) vegetation index, MODIS LST data, and ancillary data (elevation, latitude, longitude, and Julian day)(Shwetha and Kumar 2016). Duan et al., proposed a framework for the retrieval of all-weather LST at moderate spatial resolution by combining the advantages of TIR and PMW measurements(Duan et al. 2017). They used PMW and TIR LST data collected by AMSR-E and MODIS sensors aboard the Aqua satellite as input for an all-weather LST retrieval algorithm. Kou et al and Xu et al adopted the Bayesian maximum entropy (BME) to blend MODIS LST and AMSR-E LST over different land covers (LCs) and terrains (Tibetan Plateau and Heihe River Basin) for nighttime or/and day time(Kou et al. 2016; Xu et al. 2019). Zhang et al. proposed a new practical method to merge TIR and MW observations based on a decomposition of LST in the temporal dimension(Zhang et al. 2019). They decomposed LST into three temporal components: an annual temperature cycle component, a diurnal temperature cycle component prescribed by solar geometry, and a weather temperature component driven by weather change (Zhang et al. 2019).

However, estimating LSTs from PMW is challenging since the microwave signal varies significantly with land surface properties (e.g. soil moisture and vegetation cover) and, therefore, requires that the spatial and temporal variability of microwave emissivity is accounted for (Prigent et al. 2016). Moreover, (sub-)surface temperature retrieved from PMW measurements differs from LST retrieved from TIR measurements (skin temperature)(Zhou et al. 2017) and needs to be converted to skin temperatures (Moncet et al. 2011). The spatial resolution of PMW measurements (e.g., 25km for AMSR-E) is much coarser than that of TIR measurements. Also differences in spatial resolution need to be considered when fusing TIR and PMW LST products. Thus, PMW measurements are commonly regarded as complementary information to the available TIR LST (e.g. MODIS) or other auxiliary data when retrieving spatially complete LSTs at high spatial resolution (Duan et al. 2017; Shwetha and Kumar 2016; Xu et al. 2019; Zhang et al. 2019).

*3.2.2 SEB-based methods*

Unlike the aforementioned methods that are directly based on remote sensing data, SEB-based methods utilize models of the physical processes controlling the various surface parameters. Jin (2000) proposed a neighboring-pixel (NP) approach based on the surface energy balance to reconstruct cloudy pixel LST (Jin 2000). In this approach, cloudy pixels LSTs are interpolated from neighboring clear pixels and auxiliary in-situ data (e.g. net solar radiation, net longwave radiation, and latent heat flux). Based on Jin's approach, a temporal NP method was developed to estimate cloudy pixel LSTs from MSG/SEVIRI data (Lu et al. 2011). Yu et al. proposed a spatiotemporal NP method to reconstruct cloud contaminated pixels in daily MODIS LST products(Yu et al. 2014). To reduce the strong dependence on ground-based measurements, a two-step framework was developed for reconstructing satellite-based LSTs contaminated by clouds(Zeng et al. 2018). In order to improve the reconstruction accuracy over significant reliefs or topographically complex regions, an effective method based on the land energy balance theory with full consideration of topographic factors was developed to reconstruct cloudy pixel LSTs (Yu et al. 2019). Furthermore, a revised NP method with fewer parameters was proposed to fill LST gaps under cloudy-sky condition(Yang et al. 2019). In summary, considering the surface energy balance allows NP-based methods to reconstruct the actual LST under cloudy conditions. However, these methods make some assumptions and/or require that the meteorological conditions are known, so that LST differences between clear sky and cloudy pixels can be calculated, which also introduces additional uncertainty.

Recently, Fu et al. proposed another physical model-based method for retrieving urban land surface temperatures under cloudy conditions (Fu et al. 2019). They identified two main problems of SEB- and PMW-based techniques for estimating LSTs at high spatial resolution in urban areas: 1) a lack of effective downscaling techniques for PMW LSTs due to surface heterogeneity; 2) the complicated parameterization of the surface energy balance. In order to solve these two problems, they synergistically used the coupled Weather Research and Forecasting Model (WRF)/UCM system and the random forest (RF) regression technique (hereafter named as WRFF) to effectively estimate LSTs under cloudy conditions in urban areas(Fu et al. 2019). However, the performance of the developed method may be sensitive to the parameters assigned in the coupled WRF/UCM model and its effectiveness for generating long time series of LST images requires further improvements.

## 4. Spatiotemporal fusion for high resolution

The most widely used spatiotemporal fusion methods are the spatial and temporal adaptive reflectance fusion model (STARFM) and its enhanced version (ESTARFM) (Gao et al. 2006; Zhu et al. 2010). Although they were originally proposed for surface reflectance, they can also be applied to other biophysical parameters, such as LST(Liu and Weng 2012). Over the last decade various spatiotemporal fusion methods for obtaining high spatiotemporal resolution LST have been proposed. Spatiotemporal LST fusion methods can be categorized into three groups: (1) weighted function-based methods; (2) unmixing-based methods; (3) hybrid methods. An example for weighted function-based spatiotemporal fusion methods is showed in Figure 7.

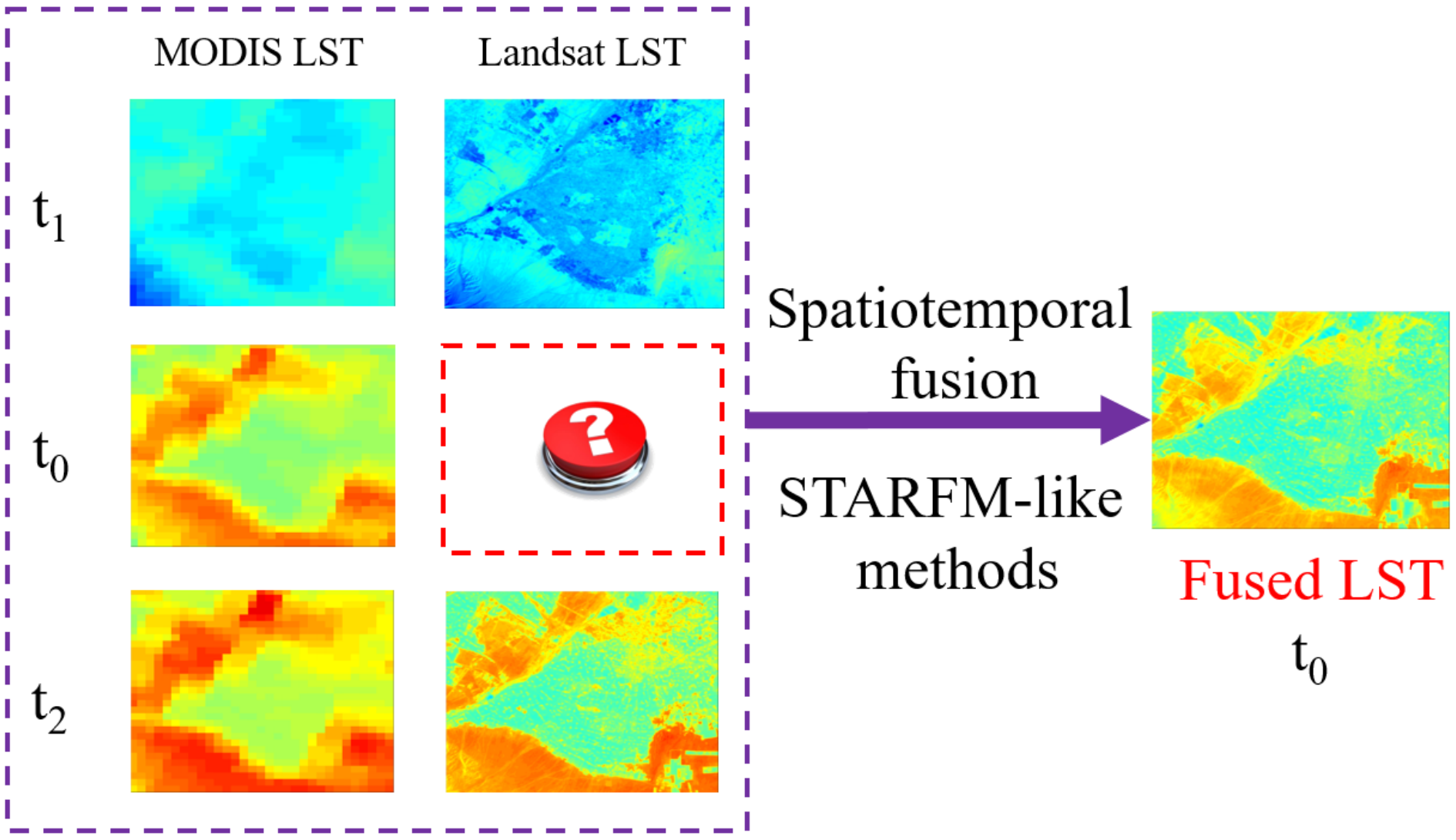

Figure 7. A diagram for weighted function-based spatiotemporal fusion methods. The two pairs of MODIS and Landsat LST images (date of $t_1$ and $t_2$) and one MODIS image at the prediction date $t_0$ can be used as the reference to compute the Landsat-scale LST image of the prediction date $t_0$.

**4.1 Weight function-based methods**

Since it was first proposed, weight function-based methods have gained considerable popularity. The most representative methods of this type are STARFM and ESTARFM. Some studies directly adopted STARFM or ESTARFM to generate daily Landsat-like thermal data, e.g. evapotranspiration and LST (Anderson et al. 2011; Li et al. 2017b; Liu and Weng 2012; Ma et al. 2018; Yang et al. 2016). STARFM-like methods with various modifications were designed for generating high spatiotemporal LST (Hazaymeh and Hassan 2015; Huang et al. 2013; Kim and Hogue 2012; Quan et al. 2018; Weng et al. 2014; Wu et al. 2013; Wu et al. 2015c): the modifications mainly concern the model relationship between different resolutions pixels, weight function design, number of sensors, and the temporal change modeling or thermal landscape representation, which aims at improving spatiotemporal LST patterns. Based on the STARFM framework, Huang et al (2013) and Wu et al.(2013) improved the weight function for producing high spatiotemporal resolution LST(Huang et al. 2013; Wu et al. 2013). Weng, Fu, and Gao (2014) and Quan et al. (2018)

modified ESTARFM to predict LST data by considering the annual temperature cycle (ATC) and urban thermal landscape heterogeneity (Quan et al. 2018; Weng et al. 2014). Wu et al. (2015) proposed a spatio-temporal integrated temperature fusion model (STITFM) for the retrieval of high temporal and spatial resolution LST from more than two sensors, including multi-scale polar-orbiting and geostationary satellite observations(Wu et al. 2015c). Weight function-based methods generally have generally shown good performances and can be improved further, e.g. by accounting for complex surface heterogeneity and land cover type change.

### 4.2 Unmixing-based methods

Unmixing-based methods regard temporal variation at a coarse spatial resolution as the mixture of the component temporal variations at a finer spatial resolution. Zhukov et al. first proposed an unmixing-based multisensor multiresolution fusion framework to integrate satellite images with different spatial resolution acquired at different times(Zhukov et al. 1999). Generally speaking, unmixing-based methods have two major challenges: large errors associated with spectral unmixing and a lack of within-class variability of the fine scale pixels inside a coarser pixel (Zhu et al. 2018b). The following unmixing-based methods are designed to address these two challenges: for instance, Zurita-Milla et al. introduced constraints into the linear unmixing process to ensure that the obtained reflectances are positive and within a physically meaningful range (Zurita-Milla et al. 2008). In order to account for land cover change, Wu et al. proposed the spatial and temporal data fusion approach (STDFA) and its modified version (ESTDFA), which use two or more image pairs with a constant and an adaptive moving window, respectively (Wu et al. 2015a; Wu et al. 2012). Huang et al. described an unmixing-based fusion method capable to account for phenological and land-cover changes(Huang and Zhang 2014). However, due to the complexity of temperature unmixing, these methods are often applied to reflectance data but not to LST data. Wu et al. estimated high spatial and temporal resolution LST for environmental process monitoring using STDFA and compared the results with STARFM and ESTARFM (Wu et al. 2015b). Their results showed that ESTARFM had the best performance, followed by STDFA and STARFM. In summary, unmixing-based algorithms generally suffer from inaccurately estimated endmember numbers, endmember spectral variability in multi-temporal observations, and spectral mixing nonlinearities; furthermore, they are challenged if land cover changes.

### 4.3 Hybrid methods

Hybrid methods integrate two or more techniques and contain weight function-based or/and unmixing-based methods. Their aim is to improve the performance of spatiotemporal data fusion through combining the advantages of different methods. Some representative methods are the Spatial and Temporal Reflectance Unmixing Model (STRUM)(Gevaert and García-Haro 2015), the Flexible Spatiotemporal Data Fusion (FSDAF)(Zhu et al. 2016), the Spatial-Temporal remotely sensed Images and land cover Maps Fusion Model (STIMFM)(Li et al. 2017a), the BLEed Spatiotemporal Temperature (BLEST)(Quan et al. 2018), and the weighted Combination of Kernel-driven and Fusion-based Methods (CKFM)(Xia et al. 2019). The first three methods combine ideas from unmixing-based methods, weighted function-based methods, Bayesian theory and spatial interpolation. Hybrid methods can fuse reflectance images under challenging scenarios, e.g. heterogeneous landscapes and abrupt land cover changes occurring between input images and prediction(Zhu et al. 2018b). Inspired by SADFAT and FSDAF, the BLEST method fuses land surface temperature of Landsat, MODIS and a geostationary satellite (FY-2F) with different spatiotemporal resolutions. BLEST adopts weight functions to better preserve spatial details; it uses

ATC and DTC models to characterize nonlinear temporal patterns; it combines a linear temperature mixing model (LTMM) and thin plate spline (TPS) downscaling to account for effects from landscape heterogeneity and land cover change. However, assuming a simple linear relationship regarding the thermal mixture over heterogeneous landscapes may not be a sufficiently good approximation of reality. Furthermore, literature shows that kernel-driven methods can obtain abundant spatial detail from visible bands, while the fusion-based process is applied for its spatiotemporal prediction ability. To fully utilize the advantages of kernel-driven methods and fusion-based methods, Xia et al. (2019) proposed a weighted combination of kernel-driven and fusion-based methods (CKFM) to enhance the resolution of LST time series (Xia et al. 2019). However, CKFM cannot be directly applied at an annual scale because the long time interval may render the regression function invalid, enlarge spatial differences, and land cover types may change (Xia et al. 2019).

Other spatiotemporal methods originally designed for reflectance images also have great potential for LST, e.g. Bayesian methods(Li et al. 2013a; Xue et al. 2017, 2019) and learning-based methods(Huang and Song 2012; Liu et al. 2019; Liu et al. 2016b; Song and Huang 2012; Song et al. 2018; Tan et al. 2018).

**5. Validation of reconstructed and fused LST products**

Because reconstruction and fusion methods are to solve the "vacancy" or "sparse" problem of satellite LST data, there is no corresponding reference LST that could be used for validation. Therefore, the validation of reconstructed and fused LST product accuracy mainly relies on simulated experimental results and cross-validation with real experimental results (Wu et al. 2015c). When simulated or real experimental results are more similar to actual LST, the reconstructed or fused LST products are to be preferred over the original degraded LST data. Cross-validation is mainly performed against ground-based, other satellite-based and data assimilation-based LST data.

It is still difficult to accurately simulate LST under cloudy sky conditions. For reconstructed LST of missing pixels and fused LST, simulation validation and cross-validation can be conducted in parallel. Based on the type of validation data, four different validation methods are distinguished:

**5.1 Simulated validation (SiV)**

Simulated validation is the most commonly used validation method. LST data from reconstruction or fusion processing are directly compared with actual LST data (Shen et al. 2016a). The closer the agreement between generated LST and actual LST, the better the performance of the reconstruction or fusion method. In practice, simulated validation is performed by first degrading the original LST image, e.g. by masking out some pixels, and then using the original LST image as reference for the reconstructed result.

**5.2 Ground-based validation (GrV)**

This approach is practicable if in situ measurements within the reconstructed or fused LST image are available. Except for a few studies that used air temperature from stations, this validation method usually involves comparisons with ground-based LST measurements, which have frequently been used to validate LST products for MODIS, GOES, SEVIRI and VIIRS(Guillevic et al. 2014; Kabsch et al. 2008a; Li et al. 2014a; Li et al. 2013b; Sun and Pinker 2003; Wang and Liang 2009; Yu et al. 2012). Usually, the in situ LSTs were used to directly evaluate the predicted LST (or spatial averages) obtained from fusion or reconstruction (Wu et al. 2013; Wu et al. 2015c; Yu et al. 2019; Zeng et al. 2018). Note that this approach can be used to validate reconstructed LST of cloudy pixels.

The three main limitations of this approach are regional restriction, spatial representativeness of the in situ reference measurements and directional effects(Guillevic et al. 2017). Although existing in situ networks for LST reference measurements are commonly used to validate current standard and research LST products, e.g. NOAA’s Surface Radiation (SURFRAD) network(Wang and Liang 2009) and KIT’s permanent validation stations operated within the framework of EUMETSAT’s Land Surface Analysis Satellite Application Facility (LSA SAF)(Göttsche et al. 2016; Gottsche et al. 2013; Kabsch et al. 2008b), in situ reference measurements are still sparse. Ground-based validation can only be performed if in situ reference measurements are located within the reconstructed and fused LST images. Furthermore, LST measured by a station does not necessarily represent a coarser satellite sensor footprint (Guillevic et al. 2017). Finally, most field radiometers collect observations near nadir, whereas wide field-of-view satellite scanners like MODIS and VIIRS collect observations from nadir to around 60° view angle(Duan et al. 2019). Therefore, in situ reference data are selected based on the following criteria: networks with high quality instrumentation and maintenance and good spatial representativeness for the satellite sensor footprint(Guillevic et al. 2017).

### 5.3 Other satellite-based validation (OsV)

This approach involves comparing reconstructed or fused LST products with heritage LST products. The method is particularly valuable when no in situ reference measurements within the reconstructed or fused LST image are available. Moreover, this approach can also be used to validate the reconstructed LST of cloudy pixels, e.g. when LST for a microwave sensor are available.

However, the approach does not yield absolute validation results and satellite LST inter-comparisons alone are insufficient to validate reconstructed or fused LST products(Guillevic et al. 2017), i.e. different retrieval algorithms based on similar assumptions and formulations (e.g. split-window) can be highly consistent with each other but biased when compared to ground reference LST(Guillevic et al. 2014).

### 5.4 Data assimilation-based validation (DsV)

This approach involves comparing reconstructed or fused LST products with LST from land data assimilation system(Rodell et al. 2004). The method is similar to other satellite-based LST validation (OsV) and can be used to validate reconstructed or fused LST products. Studies show that data assimilation-based LST products are frequently in good agreement with actual observations. However, the low spatial resolution of the data assimilation-based LST products can cause errors when validating reconstructed or fused LST products at high spatial resolution.

### 5.5 Quantitative evaluation indices (QEI)

It is important to quantitatively evaluate the accuracy of reconstruction and fusion results. A number of quantitative indicators for evaluating reconstruction and fusion results were proposed (Guillevic et al. 2017; Herrero-Huerta et al. 2019; Shen et al. 2015; Wu et al. 2015c) . The main evaluation indicators include Root mean squared error (RMSE), Mean Squared Error (MSE), Mean Absolute Error (MAE) or average absolute difference (AAD), Mean Error (ME), correlation coefficient (CC), and standard deviation (SD).

Table 2 Validation methods and quantitative evaluation indicators. (√ is feasible , √ √ is recommended and × is inapposite)

| | SiV | GrV | OsV | DsV | Quantitative evaluation indicator |
|---|---|---|---|---|---|
| Reconstruction of cloudy pixels | × | √ √ | √ √ | √ | RMSE、MSE、MAE or AAD、ME、CC or SD |
| Reconstruction of missing pixels | √ √ | √ | √ | √ | |
| Spatiotemporal fusion | √ √ | √ | √ | √ | |

## 6. Future prospects

Despite of the above achievements, the study of obtaining spatially continuous and high resolution LST (SCHR-LST) is currently one of the hot topics. However, LST reconstruction and spatiotemporal fusion are complicated and inherently ill-posed inverse problems and there still exists great space for further development. Here we list and discuss several research topics that appear promising for improving SCHR-LST retrieval from space-based measurements.

### 6.1 Exploitation of geographical laws and signal processing techniques.

Geographic surface parameters (e.g. LST) occur in the realm of space and time and, thus, follow geographical laws describing the behavior of their spatiotemporal autocorrelation, heterogeneity and spatial similarity. The first two geographical laws state that 1) geographic environmental variables are spatiotemporally correlated with themselves and 2) vary in space and time. The recently proposed Third Law of Geography focuses on the similarity of geographic configuration of locations(Zhu et al. 2018a). Spatial prediction of geographic environmental variables can be made on the basis of the similarity of geographic configurations between a sample and a prediction point. However, signal processing techniques are usually employed for establishing the global numeric relationships between variables without or with insufficient consideration of geographical laws.

Nearly all current reconstruction and spatiotemporal fusion methods originate from signal processing. There methods perform well over small and homogeneous geographic areas without significant land use/cover changes. However, when facing large and complex geographic areas, auxiliary geographical information should be utilized. LST is a typical geographical parameter and, thus, an integration of geographical laws and signal processing techniques into SCHR-LST algorithms appears to be promising.

**6.2 Integration of physical properties and signal processing techniques.**

Although input multi-temporal and/or multi-sensor LST products are retrieved with physical models, obtaining reconstructed and fused LST often involves filtering, interpolation, regression, variational processing and sparse representation methods. Previous studies showed that the accuracy of LST reconstruction significantly differs between nighttime and daytime and varies with season (Pede and Mountrakis 2018). Physical properties, such as high dynamic change characteristics, DTC and ATC, might explain the above results better than individual LST values.

Furthermore, almost all reconstruction and spatiotemporal fusion methods are implemented on LST images from multi-temporal and/or multi-sensor. However, as reviewed by Li et al., LST varies with viewing zenith angle (VZA) and acquisition time (local solar time)(Li et al. 2013b). Differences in LST measured at nadir and off-nadir can be up to 5 K for bare soils and may reach 10 K for urban areas(Li et al. 2013b). Because most polar-orbiting satellites scan the land surface in cross-track direction, their VZAs vary from –65° to +65° and can make the LSTs of different pixels in the same orbit incomparable. This effect must also be considered for LST products obtained from different sensors or at different times. In addition, LST products derived from the same satellite cannot be compared if their difference in local solar observation time is significant(Duan et al. 2014; Zhao et al. 2019). This phenomenon also affects LST products acquired by different satellites at different times. Inevitably, angular and temporal differences pose great challenges to the reconstruction and spatiotemporal fusion. Therefore, it is necessary to integrate physical properties (i.e. DTC modeling, ATC modeling, angular and temporal normalization) into reconstruction and spatiotemporal fusion methods.

**6.3 The new processing framework of reconstruction and spatiotemporal fusion**

It is well known that LST changes rapidly in space and time (Prata et al. 1995).

Traditional signal processing framework (e.g. filtering, interpolation, regression , variational processing and sparse representation) assumes that input data and output results are linear transformations or simple nonlinear transformations. It is hard to conduct comprehensive feature mining for complex nonlinear transformation process, non-stationary characteristics (such as high dynamic change characteristics) and large scale differences between high resolution LST images and low resolution LST images.

Over recent years, deep learning has gained the attention of the remote sensing community and has been used for various image understanding and image classification problems. An overview of deep learning for data fusion is provided by Zhang et al. (Ball et al. 2017; Zhang et al. 2016). It is possible to model complex relationships between different images so that a trained model could be used to predict LST with SCHR. Among the recently developed methods are reconstruction and spatiotemporal fusion based on the deep learning framework. For instance, Malek et al. and Zhang et al. proposed effective CNN models to recover missing data in remote sensing images, respectively. Song et al., Tan et al. and Liu et al. proposed novel spatiotemporal fusion models using DCCN(Song et al. 2018), DCSTFN(Tan et al. 2018) and STFNet(Liu et al. 2019) for fusing Landsat and MODIS reflectance data from different perspectives. Despite its proven efficiency for reflectance images, deep learning has rarely been used for LST retrieval. Wu et al., proposed a multiscale feature connection GLST reconstruction CNN (MFCTR-CNN) for geostationary satellite LST images with large missing regions(Wu et al. 2019): although MFCTR-CNN was tested with geostationary satellite LST, it provides a new framework and advanced capabilities for reconstructing polar-orbiting LST or other remotely sensed data products. In the future, we also could see deep learning studies on fusing LST images with high spatiotemporal resolutions.

However, imperfect knowledge of remotely sensed LST and low computing efficiency are the key factors limiting the wide application of deep learning to reconstruction and spatiotemporal fusion. However, the available LST data will continue to grow and improve with the development of big data methods and further advances in remote sensing. In addition, more advanced programming strategies (i.e. parallel computing and GPU computing) and advanced cloud-based geospatial processing platforms (e.g. Google Earth Engine, GEE) were put forward(Chen et al. 2017; Gorelick et al. 2017), thereby addressing problems with computing efficiency. All these developments will improve the application potential of remote sensing LST data significantly, which is the ultimate goal of LST reconstruction and spatiotemporal fusion.

### 6.4 Combining data assimilation with reconstruction and spatiotemporal fusion.

Data assimilation provides continuous information on variables and at locations that are not directly available from proxy data, thereby filling gaps between sparse observational data(Goosse et al. 2010). However, input uncertainty easily leads to the accumulation of errors by the model performing the assimilation and large surface heterogeneity limits its application, e.g. over urban areas. Remote sensing data can provide accurate spatial information and do not accumulate errors over time. Using data assimilation, LST can be combined organically with land surface process models to improve the accuracy of simulated land surface temperatures in time and space, which provides a new approach for obtaining continuous land surface temperature data. With this approach a series of LST data can be simulated and possible trends identified. Such trend information provides a physical constraint on LST and may improve the accuracy of LST reconstruction and spatiotemporal fusion methods.

### 6.5 Synergies between reconstruction and spatiotemporal fusion

All reviewed LST reconstruction and spatiotemporal fusion methods have been implemented as separate independent processes. In general, input data for spatiotemporal fusion should be spatially continuous or filtered with a common cloud mask (Gao et al. 2006; Quan et al. 2018; Weng et al. 2014; Wu et al. 2015c). Unfortunately, cloud contamination, defective sensors and scanning gaps between orbits bring about numerous abnormal or missing values, which negatively affect LST spatiotemporal fusion (Shen et al. 2016a). Some application studies on spatiotemporal fusion simply ignore spatially discontinuous images or use synthetic products (e.g. 8-day LST composite products, MOD11A2) (Jiang et al. 2019; Liu et al. 2016a; Lu et al. 2019). Synthetic products are helpful for analyzing dynamic change in long time series, but they increase the uncertainty in location- and time-specific quantitative studies. LST fusion results can be used as input data for reconstruction (high resolution): therefore, from an application point of view, it is highly desirable to synergistically combine LST reconstruction and spatiotemporal fusion.

## 7. Conclusions

Missing information and data gaps are a common phenomenon in satellite-retrieved LST and there is always a tradeoff between temporal and spatial resolution. Obtaining spatially continuous and high resolution LST (SCHR-LST) is crucial to many fields of research and applications, e.g. Earth's surface water and energy balances, material and energy exchange on a global scale, and subpixel wildfire temperatures detection, urban heat island monitoring on a local or fine scale. The reconstruction and spatiotemporal fusion of LST products have become active research topics that aim at overcoming this limitation. While reconstruction methods aim at obtaining spatially continuous LST fields, i.e. filling existing data gaps, spatiotemporal fusion methods generate gap-free LST fields at high spatial and temporal resolution simultaneously. This paper reviews the recent advances achieved in reconstruction and spatiotemporal fusion of LST products, and puts forward some prospects for future study in this field. The primary contributions of this work can be summarized as the following three points.

a) Current common LST satellite products and some of their specifications are summarized. The problems of reconstruction and spatiotemporal fusion are descripted and the connections between them are discussed (Section 2).

b) A thorough overview about the current achievements in reconstruction and spatiotemporal fusion is conducted (Section 3 to Section 5). The survey covers the classifications of the existing methods, summarizes their advantages and weaknesses, and introduces some of the commonly employed validation strategies. Three kinds of reconstruction methods for missing pixels (spatial, temporal, and spatiotemporal methods), two kinds of reconstruction methods for cloudy pixels (PMW-based and SEB-based methods) and three groups of spatiotemporal fusion methods (weighted function-based, unmixing-based, and hybrid methods), four validation methods (SiV, GrV, OsV and DsV) of reconstructed and fused LST products, and main evaluation indicators are summarized.

c) Some prospects for the future study of reconstruction and spatiotemporal fusion are put forward (Section 6). The exploitation of geographical laws, Integration of physical properties, DL-based frameworks, combination of data assimilation, and synergies between reconstruction and spatiotemporal fusion are discussed, respectively.

**Acknowledgements:** The authors would like to thank NASA, USGS, NOAA, LSA SAF and the China Meteorological Administration National Satellite Meteorological Center (CMA NSMC) for providing the LST products. This research was funded by the National Natural Science Foundation of China (grant number 41501376 and 41971311) and the open fund for Discipline Construction, Institute of Physical Science, and Information Technology at Anhui University.

4076